# LegenDary 2012 Soccer 2D Simulation Team Description Paper

Pourya Saljoughi[1], Reza Ma'anijou[1], Ehsan Fouladi[1], Narges Majidi[1], Saber Yaghoobi[1], Houman Fallah[1] and Saeideh Zahedi[1]

[1] Islamic Azad University of Lahijan
Legendary2D@gmail.com

**Abstract.** In LegenDary project, we started a new research based on Agent2D in RoboCup 2D soccer simulation. In this paper, we mainly present the team algorithms and structures which we used to develop our team in separated section. We have focused on passing, dribbling and blocking skills. We improved them and made the team ready for this competition. Through pass is the most important part of our team that we work a lot on it.

## 1  Introduction

The main goals of LegenDary project are both increasing our knowledge level in artificial intelligence and helping to improve this field of study in the world. For achieving these purposes we tried so hard in last few years and improved so well as it is described in continue.

LegenDary team is competing for the first time in RoboCup 2012. The first implementation of team was based on UVA base source code [1] after a few competitions we were involved we decided to use agent 2d base source code [2] as code to work on due to the good implementation of low level layer and also by experience; it has been proven that it performs much better than UVA base. We've added our own AI methods to the code and made many changes in original code to have our own team strategies and make it more flexible and powerful than it was.

The project's research focus is also concerned about developing cooperation models for soccer playing. Although most approaches to cooperation in multi-robot systems consider the control of individual robot behavior separately from the cooperative group behavior [3], we tried to have a good cooperation in our team between different agents. Any Multi-Agent cooperation is based on how the single agent adapts to the Multi-Agent System. If every element in the system can accommodate with the system, the system is steady [ 4]. Agent's action is done based on current world model state and zone and every decision is based on main goal of team in that state and agent's goal don't conflict each other. Therefore corporation will be possible.

Our team's code consists of many different parts that are discussed in following sections. We've focused on passing, dribbling and defensive skills. Through pass and blocking system are powerful parts of our team strategy. In defensive part of team, we used both blocking system and positioning system together to have a strong defensive system and for advantage of using these abilities together we used an experiment against three different teams by using and not using blocking system.

## 2  Pass

Passing is the most important skill in every team. A good passing system can make a good scoring position and also can clear the ball from the danger area. When a player can't make a better situation for himself, and the others have a better one, the pass function will be called. For this skill we inspire from MarliK[5]. We generally can classify the passing in some categories which are mentioned below.

### 2.1  Direct Pass

A direct pass is a simple pass that its target is one of our teammate's exact positions. This skill will be used when our teammates are close enough. So when the target is far, the direct pass won't have enough accuracy and it won't arrive to the target.

First of all the appropriate target will be chosen. In choosing a good target different factors will be considered and we will rate our targets. So the summation of them will be the final result of that pass rate. A pass rate will be calculated with the formula 1. $C_n$ is the factors coefficient and $E_n$ is the value of that factor in formula 1.

$$\text{Score} = C_1 * E_1 + C_2 * E_2 + C_3 * E_3 + \ldots . \tag{1}$$

After sorting targets by their rate, the best target will be chosen and the player will kick the ball with an appropriate angle and speed to that target.

### 2.2  Through Pass

If there is no possibility to dribble or direct pass, through pass is used as a strategy to break the opponent's defense line. This skill is useful in offensive situations when passing and dribbling are not effective enough to make an attacking situation.

To see if there is a possibility to pass, we use prediction system for different points around each player to find possible points for passing. The prediction system that we used is similar to Ball Interception Model in Bahia2D team [6].The algorithm for this process is shown in table 1.

**Table 1. A simplified algorithm for through pass**

*Initialize the environment's current situations.*
*Repeat for every teammate in offensive situation:*
   *Initialize canReachBall to false*
   *Initialize r to 0*
   *Initialize t to 35*
   *While t>3*
      *While r < 5*
         *Initialize distancePoint with angle of t and radius of r*
         *Initialize targetPoint array and its members according to:*
         *targetPoint[i] ← teammate player position + distancePoint*
         *SORT targetPoint array from longest distance to shortest distance*
         *Repeat for every member of targetPoint array:*
            *Call predict to see if teammate is first player who reach the ball and SET*
          *canReachBall.*
          *If canReachBall THEN*
             *kick the ball to this reliablePoint*
             *RETURN*
        *SET r TO r+1*
      *SET t TO t-1*
   *Until next member of targetPoint is not checked.*
*Until next seen teammate is not checked and targetPoint is not reliable.*
*RETURN*

For each point, prediction algorithm considers the possibility of getting the ball to the teammate. Different factors are used. We consider ball direction, ball speed and other parameters. These points are usually closest points to the goal with different angels and different distances from each player.

To send more through pass and make better situations, it's better for teammates to be near the offside line. Therefore they can reach faster than other player to the ball. This method is implemented separately for each player who is in offensive situation and each player tries to keep a minimum distance to offside line. So more through passes happen. An example of through pass is shown bellow.

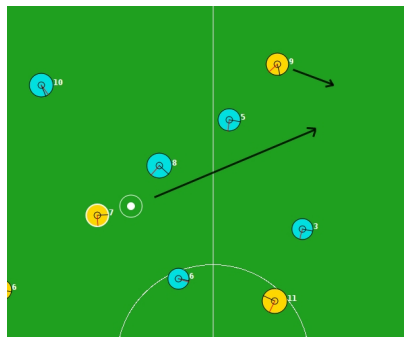

**Fig. 1.** Player 7 makes a good through Pass to player 9 which is positioned near the offside line in order to reach the ball faster.

## 3   Dribble

Dribbling is one of the most useful abilities of a team. We get the idea for this skill from ESKILAS [10]. Moving ball forward and keeping it away from the opponents is the main part of this skill. In the algorithm which we used, our main purpose is to reach the opponent's goal in predetermined ways as fast as possible. As a matter of fact paths which lead us closer to opponent's goal are in priority. When agent has the ball, this algorithm specifies which direction can place him in safer and better situation.

This process will be done with aid of player's individual vision and information about teammates' and opponents' position in the last cycles and his position towards the opponent's goal. With this information the predetermined ways will be rated and the best way will be chosen.

In our dribble skill, the dribble path is based on the predetermined paths (figure 2). So the player would rather to choose the path which is near by the main dribble path. And if he (or his chosen path) got blocked by the opponent(s), according to the position of the ball and the player itself, he decides to change the path or pass the ball to the safest teammate. If the player decides to change the path, the new path will be the nearest one to the main path and the furthest to the opponents, to keep the ball safe.

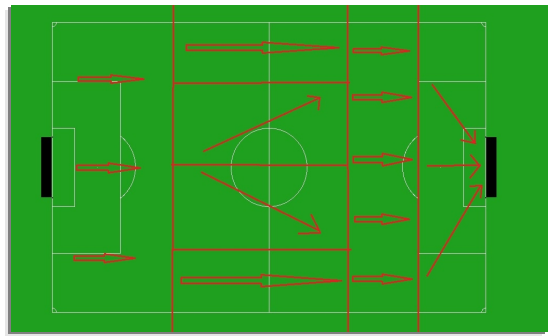

**Fig. 2.** Predetermined Paths for Dribbling

## 4   Block and Defensive system

Our Defensive system is divided into two parts. The first part is positioning. We realized that with an optimized formation system and a simple intercept action for the ball we can make base defense more powerful. Player's position is based on ball position in each cycle. We created our formation by fedit software that is available for agent2d base [7].The second part is blocking system.

Block system is one of the most important and efficient way for defending. We inspire this skill from LEAKIN'DROPS [8].

For capturing the opponent's ball, the player can choose between two different ways. The first and simple way is forcing straight ahead to the ball's position and intercepts it. The other way is to predict the target of opponent and get to the position to block the player and intercept the ball. It seems to be more effective and faster, also this method saves the stamina.

The second way is called "blocking system". As a matter of fact the first method is useless, because the opponent's position may change every moment and the player is not fixed. And the better way is forecasting the opponent's target and move through that direction to face the opponent. This method is better than the first one, because the player is not going to get dribbled and he saves time for other teammates to arrive to defense position.

Our blocking algorithm calculates a point (which is the best) between the opponent's direction and our goal according to the information he receives from the servers, like position, velocity and direction of the ball. According to the calculations, if the chosen point wasn't too far to approach and the agent was not so weak on stamina, the blocking commands will be executed.

Our blocking system is somehow dynamic and intelligent. Its decision changes according to the zone and opponent goal. We defined two ways to intercept the ball after standing in the proper block position. The player decides whether standing right there and waits for the opponent (with ball) to get close, or forcing the opponent and intercepts the ball.

We made an experiment to test our blocking system. In this experiment, we use 3 teams in 3 level of attacking power, Oxsy [9], MarliK, Helios Base. We selected these teams according to average scored in RoboCup 2011. We run 10 games against each teams (5 with blocking system and 5 without blocking system). The results of these matches are mentioned in table 2.

**Table 2. Experiment's results**

| Team Name | Blocking System Condition | M1 | M2 | M3 | M4 | M5 | Total score | Average Scored | opponent average score in RoboCup 2011 |
|---|---|---|---|---|---|---|---|---|---|
| Oxsy | on | 1-2 | 0-0 | 1-0 | 1-0 | 1-0 | 4:2 | 0.8 | 8.5 |
|  | off | 3-0 | 2-0 | 6-0 | 6-0 | 3-0 | 20:0 | 4 |  |
| MarliK | on | 0-0 | 2-0 | 0-0 | 1-0 | 1-1 | 4:1 | 0.8 | 3.92 |
|  | off | 3-0 | 3-0 | 4-0 | 2-0 | 3-0 | 15:0 | 3 |  |
| Helios Base | on | 1-6 | 0-6 | 1-4 | 0-3 | 1-4 | 3:23 | 0.6 | - |
|  | off | 2-3 | 6-3 | 6-3 | 3-4 | 4-6 | 21:19 | 4.2 |  |

As we see in the table, we can decrease the average against scored from 3.73 to 0.73 per match with blocking system. And this result shows that our blocking system is useful.

## 5  Future Work

In this paper we described some parts of our team. We spent a lot of time to improve our team's way of playing in different situation. But we have to work more and more to achieve our goal which we mentioned in the introduction section and there are several problems which must be corrected and some bugs that need to be fixed. In the next phase of LegenDary project we are going to improve our team and make it more stable and realistic by using AI algorithms.